\documentclass[numbib]{ptephy_v2}

\preprintnumber{XXXX-XXXX} 
\usepackage{hyperref}

\usepackage{tablefootnote}
\usepackage{threeparttable}

\begin{document}

\title{First Extraction of the Matter Radius of $^{132}$Sn via Proton Elastic Scattering at 200\,MeV/Nucleon}


\author[1,2,3]{Y.~Hijikata$^{*,}$}
\author[2,1]{J.~Zenihiro}
\author[4,5]{S.~Terashima}
\author[5,6,7]{Y.~Matsuda}
\author[5]{H.~Sakaguchi}
\author[8,9,10]{P.~Arthuis}
\author[11]{T.~Miyagi}
\author[5,12]{S.~Ota}
\author[1]{H.~Baba}
\author[13]{S.~Chebotaryov}
\author[2]{M.~Dozono}
\author[14,1]{T.~Harada}
\author[15]{T.~Furuno}
\author[7]{C.~Iwamoto}
\author[16]{T.~Kawabata}
\author[12]{M.~Kobayashi}
\author[17]{A.J.~Krasznahorkay}
\author[18]{S.~Leblond}
\author[18]{T.~Lokotko}
\author[19]{Y.~Maeda}
\author[12]{S.~Masuoka}
\author[12]{M.~Matsushita}
\author[1]{S.~Michimasa}
\author[1]{E.~Milman}
\author[1,2]{T.~Murakami}
\author[7]{H.~Nasu}
\author[7]{J.~Okamoto}
\author[20]{S.~Sakaguchi}
\author[12]{M.~Takaki}
\author[19]{K.~Taniue}
\author[12]{H.~Tokieda}
\author[2]{M.~Tsumura}
\author[21]{O.~Wieland}
\author[22]{Z.H.~Yang}
\author[12]{Y.~Yamaguchi}
\author[12]{R.~Yokoyama}
\author[1,3]{T.~Uesaka}

\affil[1]{RIKEN Nishina Center for Accelerator-Based Science, Wako, Saitama 351-0198, Japan \email{yuto.hijikata@riken.jp}}
\affil[2]{Department of Physics, Kyoto University, Sakyo, Kyoto 606-8502, Japan}
\affil[3]{RIKEN Cluster for Pioneering Research, Hirosawa, Wako, Saitama 351-0198, Japan}
\affil[4]{Institute of Modern Physics, Chinese Academy of Sciences, Lanzhou 730000, China}
\affil[5]{Research Center for Nuclear Physics, University of Osaka, Ibaraki, Osaka 567-0047, Japan}
\affil[6]{Department of Physics, Konan University, Hyogo 658-8501, Japan}
\affil[7]{Research Center for Accelerator and Radioisotope Science, Tohoku University, Sendai, Miyagi 980-8578, Japan}
\affil[8]{Université Paris-Saclay, CNRS/IN2P3, IJCLab, 91405 Orsay, France}
\affil[9]{Technische Universität Darmstadt, Department of Physics, 64289 Darmstadt, Germany}
\affil[10]{ExtreMe Matter Institute EMMI and Helmholtz Forschungsakademie Hessen für FAIR (HFHF), GSI Helmholtzzentrum für Schwerionenforschung GmbH, 64291 Darmstadt, Germany}
\affil[11]{Center for Computational Sciences, University of Tsukuba, Tsukuba, Ibaraki, Japan}
\affil[12]{Center for Nuclear Study, the University of Tokyo, Wako, Saitama 351-0198, Japan}
\affil[13]{Department of Physics, Kyungpook National University, Daegu 41566, Republic of Korea}
\affil[14]{Department of Physics, Toho University, Funabashi, Chiba 274-8510, Japan}
\affil[15]{Department of Applied Physics, University of Fukui, Fukui, Fukui 910-8507, Japan}
\affil[16]{Department of Physics, University of Osaka, Toyonaka, Osaka 560-0043, Japan}
\affil[17]{HUN-REN Institute for Nuclear Research (HUN-REN ATOMKI), P.O. Box 51, H-4001 Debrecen, Hungary}
\affil[18]{Department of Physics, The University of Hong Kong, Hong Kong, China}
\affil[19]{Faculty of Engineering, University of Miyazaki, Miyazaki 889-2192, Japan}
\affil[20]{Department of Physics, Kyushu University, Fukuoka 819-0395, Japan}
\affil[21]{INFN Sezione di Milano, Via Celoria 16, Milano, I-20133, Italy}
\affil[22]{School of Physics and State Key Laboratory of Nuclear Physics and Technology, Peking University, Beijing 100871, China}

\begin{abstract}%
The angular distribution of the differential cross sections for proton elastic scattering from $^{132}$Sn at 196--210\,MeV/nucleon was successfully measured over a momentum transfer range of 0.80 to 2.1\,fm$^{-1}$.
Using a relativistic impulse approximation, the root-mean-square matter radius of $^{132}$Sn was extracted to be $4.758^{+0.023}_{-0.024}$\,fm, which was compared with the state-of-the-art \textit{ab initio} calculations.
Combined with the charge radius measured at ISOLDE, there are no theoretical calculations consistent with both matter and charge radii within the experimental errors.
\end{abstract}

\subjectindex{xxxx, xxx}

\maketitle

\section{Introduction}
The size of a nucleus is one of the most fundamental properties, characterizing the many-body system composed of protons and neutrons.
Recent advances in accelerator technologies and radioactive-isotope (RI) production methods have enabled us to access unstable neutron-rich nuclei with large isospin asymmetry, drawing attention to their radii.
Among the radii of nuclei accessible at the existing facilities, that of $^{132}$Sn is of particular significance and attracts keen interest.
The $^{132}$Sn nucleus is one of a few double-magic nuclei located far from the stability line and has a large isospin asymmetry of $(N-Z)/A = 0.242$.
The asymmetry exceeds that of $^{208}$Pb ($(N-Z)/A=0.221$), whose neutron-skin thickness, the difference between the root mean-square (rms) radii of neutrons and protons, has been regarded as the de facto standard in the studies of the symmetry energy term of the nuclear equation of state (EOS)~\cite{Trzcinska2001, Zenihiro2010, Tamii2011, Tarbert2014, Adhikari2021}.
Because of its larger isospin asymmetry, the radii of $^{132}$Sn are expected to impose more stringent constraints on the symmetry energy parameters than those of $^{208}$Pb~\cite{Chen2005}.
Moreover, thanks to remarkable progress in theoretical approaches and computational technologies, \textit{ab initio} calculations starting from chiral effective field theory (EFT) are now applicable to heavy nuclei, including $^{132}$Sn~\cite{Arthuis2020, Miyagi2022, Hu2022, Arthuis2024}.
As a result, measurements of nuclear radii allow for direct examinations of nuclear forces derived from chiral EFT.

Laser spectroscopy and electron scattering have been used to determine charge radii of stable and unstable nuclei~\cite{Angeli2013}.
Actually, the charge radius of $^{132}$Sn has been measured as 4.7093(76)\,fm using the laser spectroscopy at ISOLDE~\cite{Blanc2005, Gorges2019}.
Recently, the first electron scattering experiment on unstable nuclei, $^{137}$Cs, was conducted by the Self-Confining RI Ion Target (SCRIT) project at RIKEN \cite{Tsukada2023}.
Upgrades of SCRIT are ongoing toward the electron scattering of the $^{132}$Sn nucleus.

Accurate determination of the matter radius, or the neutron radius, is experimentally more demanding.
Since the discovery of the halo structure in $^{11}$Li~\cite{Tanihata1985}, the reaction cross section for a nuclear target is known to be efficient in determining the matter radii of light neutron-rich nuclei~\cite{Ozawa2001}.
Recently, the method has been applied to neutron-rich calcium isotopes~\cite{Tanaka2020} and to the tin isotopes~\cite{Kudaibergenova2024}.
Although the parity-violating electron scattering is regarded as the most accurate experimental approach to the neutron radii~\cite{Adhikari2021, Ahikari2022}, its measurement for unstable nuclei is far from reality at present.

Proton elastic scattering at the intermediate energy region is a promising approach for determining the matter radii of unstable nuclei~\cite{Sakaguchi2017}.
The method of the proton scattering experiment was applied to light neutron-rich nuclei: In the measurements at GSI, the active target system IKAR with a hydrogen gas was introduced for the inverse-kinematics measurement, and matter radius and distributions were determined for He~\cite{Neumaier2002}, Li~\cite{Dobrovolsky2006}, Be~\cite{Ilieva2012}, B~\cite{Korolev2018}, and C~\cite{Dobrovosky2021} isotopes.
Later, the EXL collaboration conducted the proton elastic scattering experiment in the heavy-ion storage ring ESR, using the internal hydrogen gas-jet target~\cite{Schmid2023}.
The two measurements at GSI realized access to the small momentum transfer region by employing gas or gas-jet targets.

The other approach is to use solid targets for better statistics~\cite{Matsuda2013, Chebotayov2018}, which allows us to measure the cross section up to the large momentum-transfer region.
Covering such a wide momentum-transfer range is essential for determining the density distribution and obtaining the precise radius, as demonstrated in normal-kinematics measurement~\cite{Terashima2008, Zenihiro2010}.
In addition, measurements with high angular resolution are also required for the accurate determination of nuclear structure.
Therefore, we have launched the Elastic Scattering of Protons with RI beams (ESPRI) project and developed a recoil proton spectrometer (RPS)~\cite{Matsuda2013} to measure the proton elastic scattering in inverse kinematics covering a large angular region with high precision.
Using the solid hydrogen target (SHT)~\cite{Matsuda2011} and RPS, we have successfully performed the proton elastic scattering measurement from $^{9}$C~\cite{Matsuda2013} and $^{6}$He~\cite{Chebotayov2018}.

In this article, we report the first measurement of proton elastic scattering from $^{132}$Sn at 200\,MeV, covering a momentum transfer range from 0.8 to 2.1 fm$^{-1}$ using RPS.
The matter radius of $^{132}$Sn was extracted from the measured angular distribution of the cross sections using relativistic impulse approximation (RIA) calculations for the first time.
The results were also compared with the \textit{ab initio} in-medium similarity renormalization group (IMSRG) calculations~\cite{Hergert2016} using the state-of-the-art chiral EFT interactions.

\section{Experiment}
\begin{figure}[!h]
    \centering
    \includegraphics[width=1.0\linewidth]{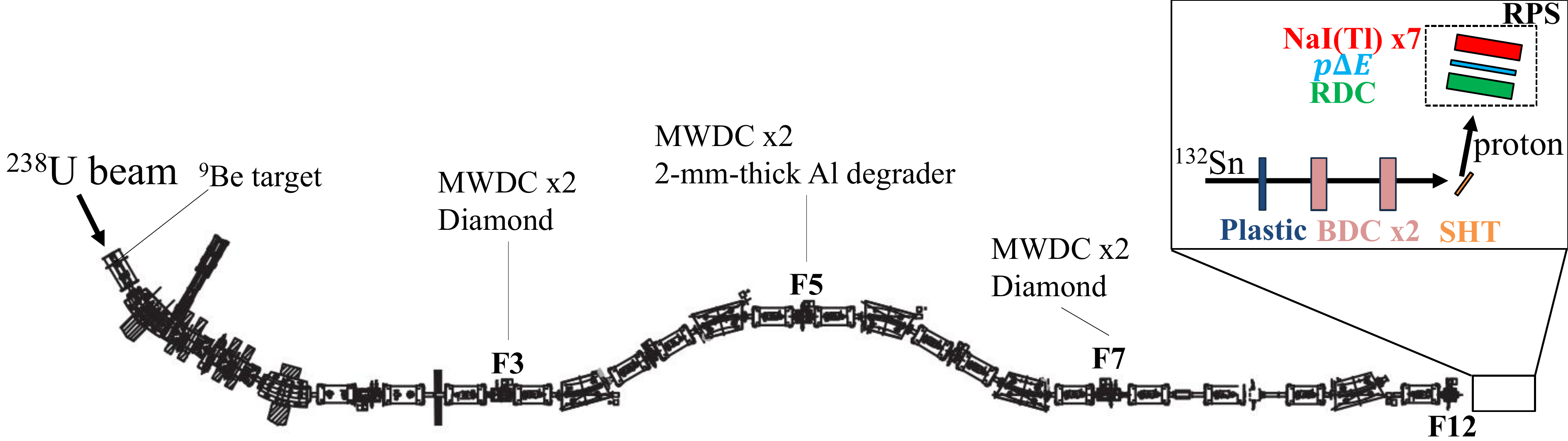}
    \caption{
    Schematic view of the beamline, detectors, and materials.
    }
    \label{fig::beamline}
\end{figure}

The experiment was performed at the RI Beam Factory (RIBF) at RIKEN.
Figure~\ref{fig::beamline} shows an overview of the beamline for the secondary beam.
A primary $^{238}$U beam at 345\,MeV/nucleon was bombarded on a 4-mm-thick $^{9}$Be target.
The secondary $^{132}$Sn beam was separated and identified using the Superconducting Radioactive Isotope Beam Separator (BigRIPS) fragment separator~\cite{Fukuda2013}.
The total beam intensity of the secondary beam was 200\,kcps at the seventh focus (F7).
A conventional particle identification (PID) method with the standard BigRIPS detectors, namely, plastic scintillators, delay-line parallel-plate avalanche counters~\cite{Kumagai2001}, and an ion chamber, is not suitable for such high-intensity heavy ion beams due to radiation damage and degraded efficiency.
Therefore, a new PID method was employed using diamond detectors~\cite{Michimasa2013}, low-pressure multiwire drift chambers (MWDCs)~\cite{Miya2013}, and a 2-mm-thick aluminum degrader.
In this method, the time of flight (TOF) between the achromatic third focus (F3) and F7 was measured by the two diamond detectors.
The energy loss in the degrader placed at the dispersive fifth focus (F5) was indirectly determined by the differences in the magnetic rigidities, derived from the TOF and the position at F5 measured by the MWDCs.
The typical purity of $^{132}$Sn was 28\%.
The energy on the secondary target was event-by-event determined from the beam position at F5, and distributed from 196 to 210\,MeV/nucleon.

\begin{figure}[!h]
    \centering
    \includegraphics[width=0.8\linewidth]{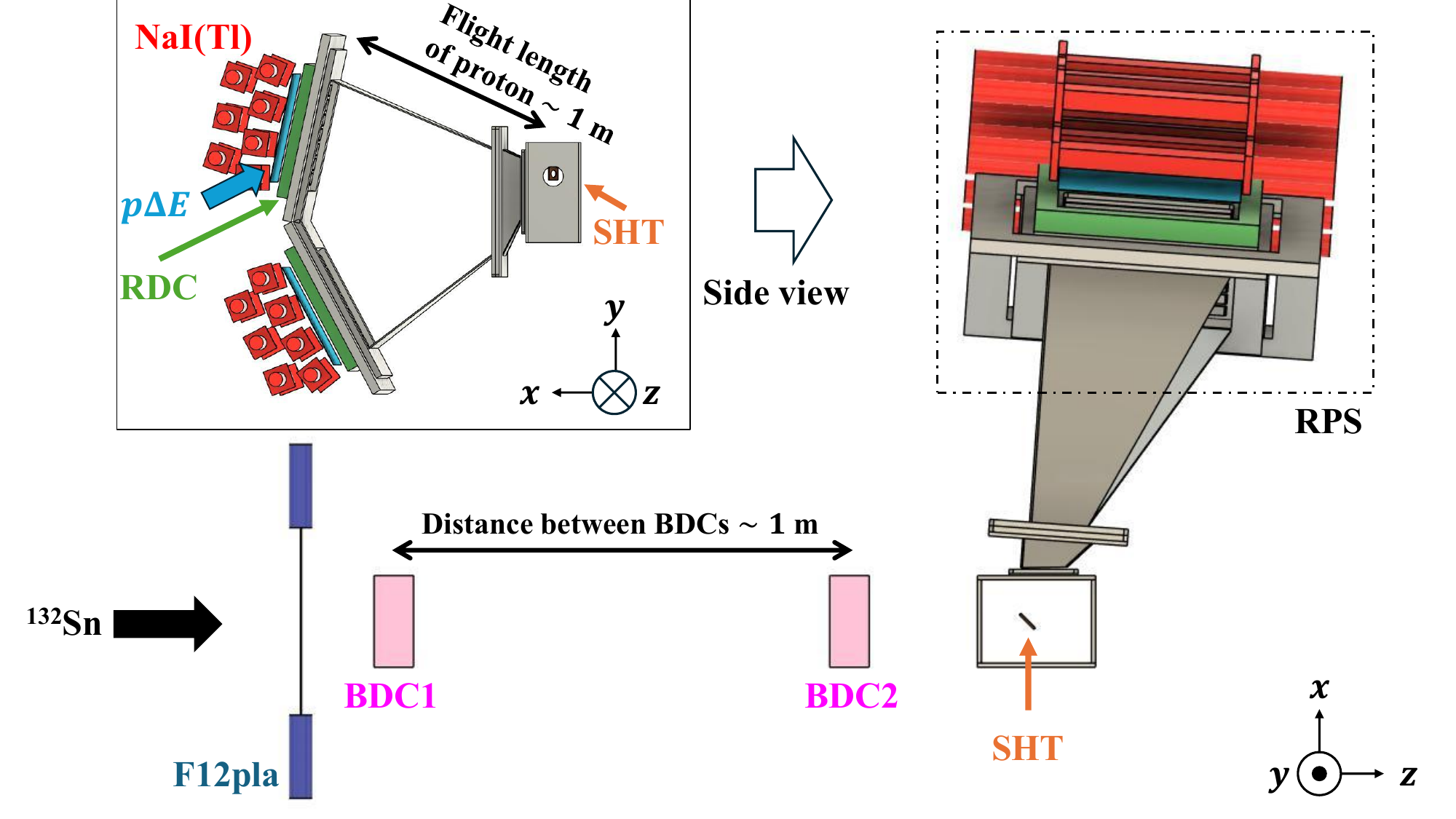}
    \caption{
    Top view of the setup around the SHT.
    The figure in the upper left shows a side view of the RPS.
    }
    \label{fig::f12setup}
\end{figure}

After passing through the BigRIPS separator, the secondary beam was transported to the twelfth focus (F12), where RPS~\cite{Matsuda2013} was installed.
Figure~\ref{fig::f12setup} shows the schematic view of the setup around the secondary target.
The plastic scintillator (F12pla) on the beamline provided timing information.
The two beamline low-pressure MWDCs (BDCs) with CH$_4$ gas, placed upstream of the solid hydrogen target (SHT)~\cite{Matsuda2011}, determined the trajectory and position of the incident beam on the SHT.
The angular resolution of the beam trajectory and the position resolution on the SHT were 0.2\,mrad and 0.2\,mm ($\sigma$), respectively.
The SHT was tilted by $45^\circ$ to reduce multiple scattering of recoil protons in the target.
It had a thickness of approximately 1.4\,mm and a diameter of 25\,mm along the beamline. 
The precise thickness was deduced by measuring the differential cross section of the proton elastic scattering from stable nuclei $^{48}$Ca at beam energies of 225--245\,MeV/nucleon during the same beam time, resulting in an evaluated number of $6.76(35)\times10^{21}$\,/cm$^2$ protons in the target.

Recoil protons from the SHT were detected using RPS~\cite{Matsuda2013}, which has two detector sets.
Each set consists of an MWDC for recoil-particle tracking (recoil drift chamber, RDC), a plastic scintillator ($p\Delta{}E$), and seven NaI(Tl) calorimeters used to reconstruct the missing-mass spectrum.
The specifications of these RPS detectors are summarized in Table~\ref{tab:RPSdet}.
Since there was a problem with the lower RDC, only the upper-side data were used for the analysis in this article.
The recoil angles were determined by combining the position on the RDC with that on the SHT by BDCs.
The position resolution of RDC was less than 0.2\,mm ($\sigma$).
The tracking efficiency of the RDC was evaluated to be about 93\%, and didn't have noticeable dependences on the position and the energy of the recoil particles.
The recoil energies of low-energy protons below 20\,MeV, such as those stopping at the $p\Delta{}E$, were determined by the TOF from the F12pla to the $p\Delta{}E$, while those of high-energy protons above 20\,MeV, such as those penetrating the $p\Delta{}E$, were determined by the NaI(Tl) calorimeters.
Protons with the energies above 120\,MeV can penetrate the NaI(Tl) rods.

\renewcommand{\arraystretch}{1.2}
\begin{table}[!h]
\centering
\caption{Specification of the RPS detectors.}
\label{tab:RPSdet}
\begin{tabular}{ll}
    \hline
    \multicolumn{2}{c}{RDC} \\
    \hline
    Effective area                  & $436\times436$\,mm$^{2}$ \\
    Plane configuration$^{a}$       & X-Y-X-Y-X'-Y'-X' \\
    Cell size                       & $14\times14$\,mm$^2$ \\
    Filling gas                     & Ar($50\%$) + C$_2$H$_6$ ($50\%$) 1\,atm \\
    Cathode and Potential wires     & 30\,$\mu$m$\phi$ Au-W/Re \\
    Anode wire                      & 100\,$\mu$m$\phi$ Be-Cu \\
    Cathod and Pontential voltages  & $-2.1$\,kV \\
    \hline
    \hline
    \multicolumn{2}{c}{$p\Delta{}E$ (plastic scintillator)} \\
    \hline
    Effective area & $440\times440$\,mm$^{2}$  \\
    Thickness      & 2.53\,mm (up), 3.09\,mm (down) \\
    \hline
    \hline
    \multicolumn{2}{c}{NaI(Tl)} \\
    \hline
    Effective area & $431.8\times45.72$\,mm$^{2}$ \\
    Thickness      & 50.8\,mm \\
    \hline
    \multicolumn{2}{l}{a) The wire positions of the X' and Y' planes are shifted by half cells. }\\
\end{tabular}
\end{table}
\renewcommand{\arraystretch}{1.0}

\begin{figure}[!h]
    \centering
    \includegraphics[width=0.8\linewidth]{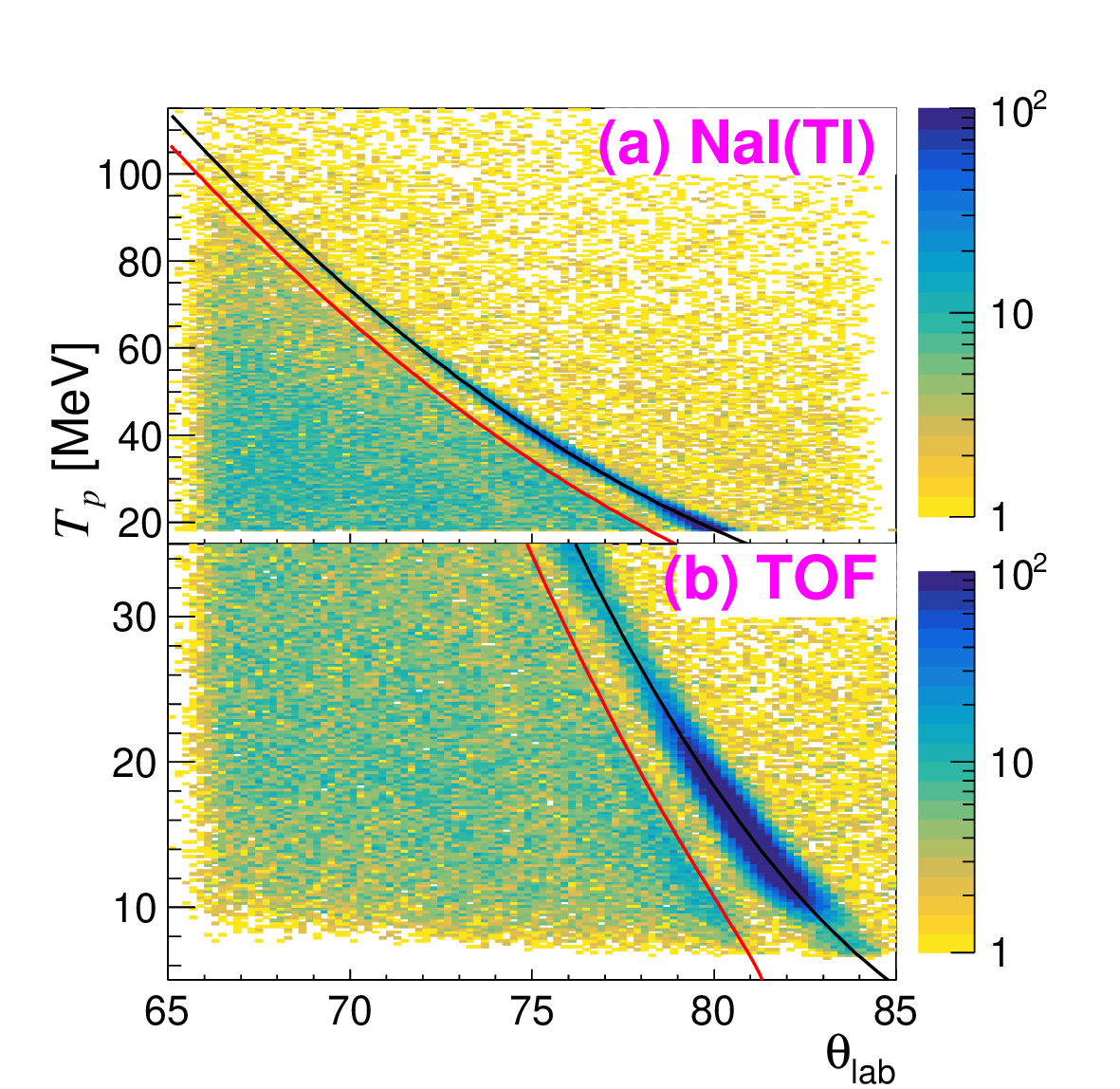}
    \caption{
    Correlations between the angle and energy for recoil protons.
    The energies are measured by the NaI(Tl) scintillators (top panel) and the TOF (bottom panel), respectively.
    The black and red lines show the kinematical correlations of the elastic scattering and inelastic scattering to the first excited state (4.04\,MeV).
    }
    \label{fig::kinema}
\end{figure}

Figure~\ref{fig::kinema} shows the 2D plot between the energies and angles of the recoil protons.
The black and red lines are the kinematical correlations of the elastic scattering and inelastic scattering with the first excited state (4.04\,MeV), respectively.

\begin{figure}[!h]
    \centering
    \includegraphics[width=0.8\linewidth]{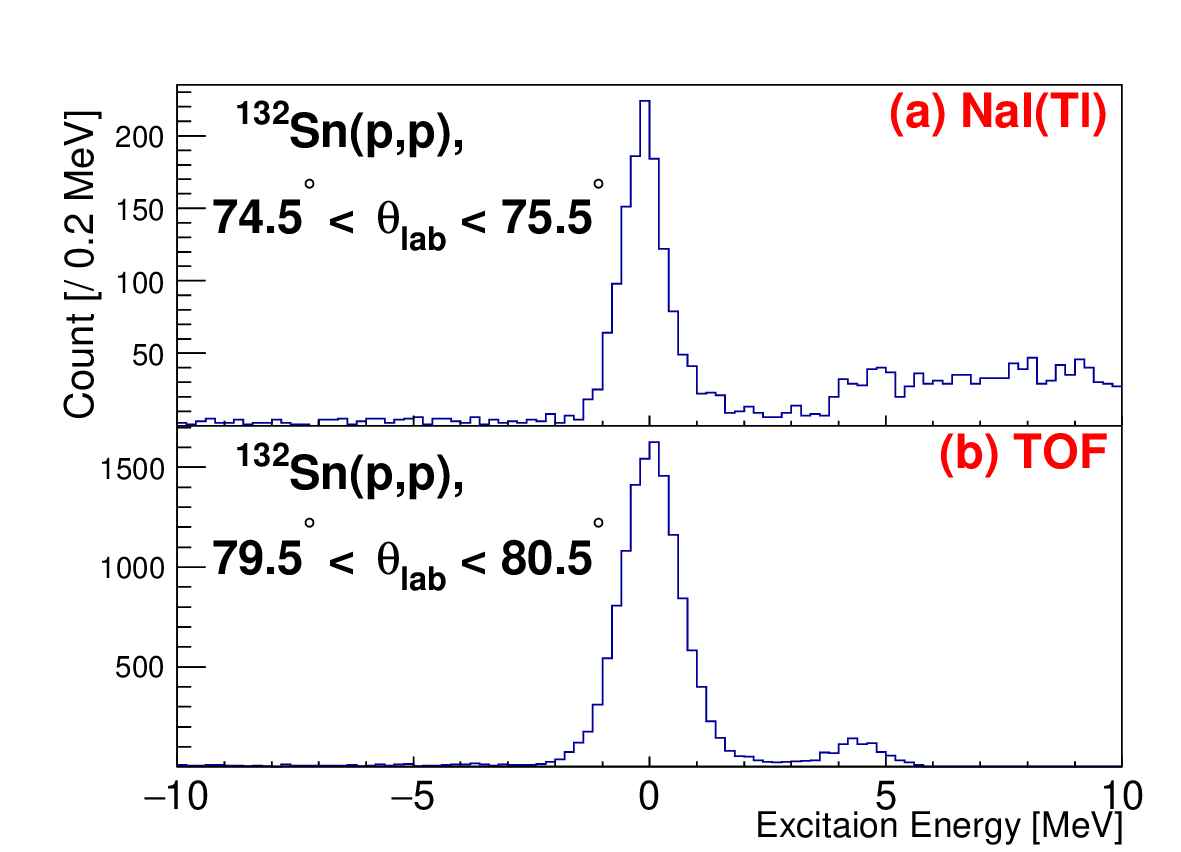}
    \caption{
    Excitation energy spectra for $^{132}$Sn(\textit{p},\textit{p}) scattering.
    The recoil proton energies are measured by the NaI(Tl) scintillators (top panel) and the TOF (bottom panel), respectively.
    }
    \label{fig::ex}
\end{figure}

By combining the obtained energies and angles of the recoil protons with those of the beams, we reconstructed the excitation energy spectrum, as shown in Fig.~\ref{fig::ex}.
In Figs.~\ref{fig::ex}(a) and \ref{fig::ex}(b), the recoil energies were determined by the NaI(Tl) calorimeters and TOF, respectively.
The typical excitation energy resolution was 0.7\,MeV~($\sigma$), allowing for a clear separation of elastic scattering from the inelastic scattering to the first excited state (4.04\,MeV).
The dominant factor in the excitation energy resolution was multiple scattering in the forward angle region, while in the backward region, the energy resolution of the NaI(Tl) calorimeters and multiple scattering contributed comparably.
The fairly flat background is found in the upper-right region of the elastic locus in Fig~\ref{fig::kinema}, and attributed to events arising from the 6-$\mathrm{\mu{}m}$-thick aramid target window.

\begin{figure}[!h]
    \centering
    \includegraphics[width=0.8\linewidth]{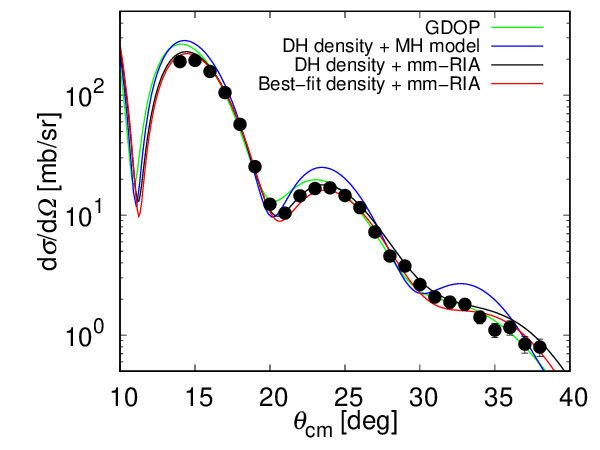}
    \caption{
    Differential cross sections~$\mathrm{d}\sigma/\mathrm{d}\Omega$ for elastic scattering from $^{132}$Sn at 196--210\,MeV/nucleon.
    The circles show the experimental result.
    The solid lines are the predictions by the GDOP (green), the MH-model with the DH density (blue), the mm-RIA with DH density (black), and the mm-RIA with the best-fit density (red).
    }
    \label{fig::CS_react}
\end{figure}

For every 1-degree step of scattering angle in the CM frame, the excitation energy spectra were fitted with a combination of a constant plus a Gaussian function.
After subtracting the constant background, events within three sigma ($3\sigma$) of the Gaussian peak were counted as elastic events.
The angular distribution of the differential cross section was obtained over the range of $14^\circ<\theta{}<38^\circ$ in the CM frame, corresponding to a range of the momentum transfer from 0.80 to 2.1\,fm$^{-1}$, as shown by the black points in Fig.~\ref{fig::CS_react}.
The main components of the errors of the cross sections in the forward and backward angular regions are the ambiguity of the SHT thickness and the statistical uncertainty, respectively.
Systematic errors other than those arising from the target number, such as the tracking efficiency of the RDC, were deduced to be about 2\%.
However, since the systematic errors related to the RPS were largely canceled in the evaluation of the target number, the actual uncertainty is expected to be smaller.

It should be noted that the scattering angles were smeared due to angular straggling, mainly caused by the multiple scattering in the SHT.
This effect is particularly significant at forward angles in the CM frame, where the recoil protons have low energies.
For instance, the angular straggling at the most forward angle of $\theta_\mathrm{cm}=14^{\circ}$ is approximately $0.3^{\circ}$ ($\sigma$), where 10\% of the events are assigned in the incorrect bins for $1^{\circ}$ angular bin width, whereas for angles larger than $\theta_\mathrm{cm}=18^{\circ}$, more than 99\% of the events are assigned in the correct bins.
Similar to the beam energy spread, the smearing of the scattering angles significantly affects the cross sections.
Therefore, when evaluating the $\chi^2$ value, we used a theoretical calculation averaged over the angular range, taking into account the smearing effect.

\section{Reaction framework}

The present results are compared with several theoretical calculations: The green line in Fig.~\ref{fig::CS_react} represents that with the global Dirac optical potential (GDOP)~\cite{Cooper1993}.
The GDOP is known to satisfactorily reproduce data for stable nuclei.
The calculation for $^{132}$Sn reproduces the present data well for the backward angles.
However, it exhibits a noticeable deviation in shape around the first and second peaks.

RIA is one of the most successful reaction models for intermediate-energy proton elastic scattering.
We employ a RIA model formulated by Murdock and Horowitz~\cite{Murdock1987}, where the transition amplitude is obtained by folding the relativistic Love-Franey (RLF) nucleon-nucleon (NN) interaction~\cite{Love1981, Horowitz1985} over the point nucleon densities.
For the proton elastic scattering of stable nuclei, this model reproduced the spin observables particularly well.
The blue line in Fig.~\ref{fig::CS_react} is a cross section calculated by the Murdock-Horowitz model (MH-model) with the Dirac-Hartree (DH) density formulated by Horowitz and Serot~\cite{Horowitz1987} for $^{132}$Sn.
It should be noted that there is no free parameter in the calculation.
Although the calculation reproduces the oscillation pattern at around the second diffraction peak ($\theta_{\mathrm{cm}}\approx24$\,deg), its amplitude deviates considerably from the data.

As in previous works~\cite{Terashima2008, Zenihiro2010, Matsuda2013}, we developed a medium-modified RIA (mm-RIA), in which a phenomenological modification is incorporated into the RLF NN interaction to take medium effects, such as Pauli blocking, into account.
The modification is the introduction of the density-dependent parameters into the coupling constants and the mass of $\sigma$ and $\omega$ mesons, which are expressed as
\begin{eqnarray}
    g^2_i, \overline{g}^2_i &&\to \frac{g^2_i}{1 + a_i\rho(r)/\rho_0}, \frac{\overline{g}^2_i}{1 + \overline{a}_i\rho(r)/\rho_0}, \\ \label{eq::me1}
    m_i, \overline{m}_i &&\to m_i\Bigg[ 1 + b_i\frac{\rho(r)}{\rho_0} \Bigg], \overline{m}_i\Bigg[ 1 + \overline{b}_i\frac{\rho(r)}{\rho_0} \Bigg], \\ \label{eq::me2}
    i &&= \sigma, \omega, \nonumber
\end{eqnarray}
where $g_i$, $\overline{g}_i$, $m_i$, and $\overline{m}_i$ are the coupling constants and masses of $\sigma$ and $\omega$ mesons for real and imaginary amplitudes, respectively.
$\rho(r)/\rho_0$ is the nucleon density divided by the normal density 0.1934\,fm$^{-3}$.
The parameters $a_i$, $\overline{a}_i$, $b_i$, and $\overline{b}_i$ for 200\,MeV were newly determined to reproduce the proton elastic scattering data from $^{58}$Ni.
Because the neutron radius of $^{58}$Ni is predicted to be almost the same as the proton radius by many mean-field calculations~\cite{Ray1978_1, Ray1978_2, Trzcinska2001, Hofmann2001, Antonov2005}, the neutron density distribution is assumed to be the same as the proton density distribution ($\rho_n(r)=(N/Z)\rho_p(r)$).
The proton density distribution of $^{58}$Ni was determined by unfolding the charge density distribution~\cite{Devris1987}.
While both scalar and vector densities are necessary for the RIA calculation, scalar densities cannot be directly extracted from experiments unless the ground-state wave functions are known.
Because the ratio of the scalar to vector density is almost constant at 0.96 as reported in Ref.~\cite{Sakaguchi1998, Terashima2008}, we adopted this assumption.
In this work, when performing the mm-RIA calculations, this assumption was used for $^{132}$Sn as well as $^{58}$Ni. 
For the previous analysis of the proton elastic scattering at 300\,MeV~\cite{Terashima2008, Zenihiro2010, Matsuda2013}, the number of parameters was reduced to four by using the same values for real and imaginary modifications ($\overline{a}_i=a_i$, $\overline{b}_i=b_i$).
On the other hand, full eight-parameter modifications were used to improve the reproducibility of the angular distributions at 200\,MeV.
The validity of the mm-RIA analysis at 200\,MeV has been confirmed, since the analysis of $^{90}$Zr data at 200\,MeV yields density and radius values consistent with those obtained at 300\,MeV~\cite{Sakaguchi2017}.

\begin{figure}[!h]
    \centering
    \includegraphics[width=0.8\linewidth]{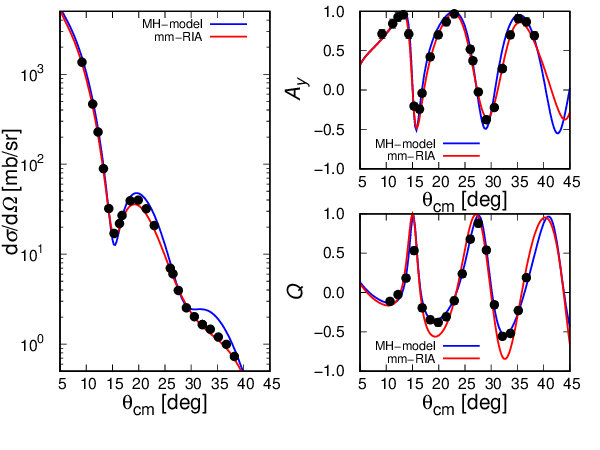}
    \caption{
    Experimental data of the cross section~$\mathrm{d}\sigma/\mathrm{d}\Omega$, analyzing power~$Ay$~\cite{Sakaguchi2017}, and spin-rotation parameter~$Q$~\cite{Takeda2003} from $^{58}$Ni at 200\,MeV.
    The red and blue lines show the best-fit mm-RIA and MH-model calculations with the deduced densities, respectively.
    }
    \label{fig::CSni58}
\end{figure}

The medium-effect parameters were calibrated by using the Monte Carlo technique to minimize the $\chi^2$ for the cross section, analyzing power, and spin-rotation parameter of $^{58}$Ni(\textit{p},\textit{p}) at 200\,MeV, which were measured at the Research Center for Nuclear Physics, University of Osaka~\cite{Takeda2003, Sakaguchi2017}.
The data up to $\theta_{\mathrm{cm}}=40^{\circ}$ corresponding to the present measurement of $^{132}$Sn were used for the fitting.
Table~\ref{tab:MEpara} summarizes the calibrated parameters.
The values of $\overline{a}_i$ are relatively larger than the parameters obtained in the previous work~\cite{Zenihiro2010}, which was based on a four-parameter fitting.
In general, the Pauli blocking effect on proton elastic scattering at intermediate energies is known to reduce the coupling constants, particularly their imaginary parts~\cite{Murdock1987}.
Therefore, the relatively large values obtained here are likely to reflect this effect.
Figure~\ref{fig::CSni58} shows the experimental data and theoretical calculations.
The mm-RIA calculation using the calibrated parameters with the deduced densities with the assumption of $\rho_n(r)=(N/Z)\rho_p(r)$ (red line) is in better agreement with the experimental data than the MH-model calculation with the DH density (blue line).

Using the mm-RIA, we calculated the cross section with the DH density for $^{132}$Sn, as shown by the black line in Fig.~\ref{fig::CS_react}.
The angular distribution better reproduces the experimental result, namely, the heights of the first and second diffraction peaks and the oscillation phase, than with the original RLF NN interactions (blue line).
In the following, the mm-RIA is employed for the extraction of the matter radius and the comparisons with the \textit{ab initio} calculations.


\begin{table}
\centering
\caption{The medium-effect parameters $a_i, \overline{a}_i, b_i$ and $\overline{b}_i$ in Eqs.~(\ref{eq::me1}) and (\ref{eq::me2}).
}
\label{tab:MEpara}
\begin{tabular}{crr}
    \hline
     $i$              & \multicolumn{1}{c}{$\sigma$} & \multicolumn{1}{c}{$\omega$} \\
     \hline
     \hline
     $a_i$            & $ 0.0210$ & $ 0.1965$  \\
     $\overline{a}_i$ & $ 2.0467$ & $ 1.1549$ \\
     $b_i$            & $ 0.0326$ & $-0.0224$ \\
     $\overline{b}_i$ & $-0.0799$ & $-0.0501$ \\
     \hline
\end{tabular}    
\end{table}

\section{Extraction of Matter Radius}
The rms matter radius of $^{132}$Sn was extracted through the following procedure.
The proton and neutron density distributions were modeled with the two-parameter Fermi (2pF) functions:
\begin{eqnarray}
    \rho_i(r) &&= \frac{n_i}{ 1 + \exp{(r-c_i)/a_i}}, \\
    i &&= p, n, \nonumber
\end{eqnarray}
where $c_i$ and $a_i$ are free parameters used to reproduce the data, and $n_i$ are normalization coefficients adjusted to conserve the number of protons and neutrons.
The matter density distribution $\rho_m(r)$ is given by the sum of the proton and neutron density distributions: $\rho_m(r)=\rho_p(r) + \rho_n(r)$.
Using the mm-RIA calculations, we searched the parameters $c_i$ and $a_i$ to minimize the $\chi^2$ value for the cross section.
The mean-square radii are calculated as
\begin{eqnarray}
    \langle{}r_{i}^2\rangle{} &&= \frac{1}{N_i}\int d\mathbf{r} \rho_{i}(r) r^2, \\
    i &&= p, n, m, ch \nonumber
\end{eqnarray}
where $N_i$ corresponds to the atomic number, neutron number, mass number, and nuclear charge for the proton, neutron, matter, and charge distributions ($\rho_p$, $\rho_n$, $\rho_m$, $\rho_{ch}$), respectively.
In the fitting, we constrained the $c_p$ and $a_p$ to reproduce the known value of the rms charge radius $\langle{}r_{ch}^2\rangle{}^{1/2}$ of 4.7093(76)\,fm, measured by the laser spectroscopy at ISOLDE~\cite{Blanc2005}.
The rms proton radius $\langle{}r_{p}^2\rangle{}^{1/2}$ calculated with $c_p$ and $a_p$ is related to the charge radius as 
\begin{equation} \label{eq::prad}
    \langle{}r_{p}^2\rangle{} \simeq \langle{}r_{ch}^2\rangle{} - \langle{}R_{p}^2\rangle{} - \frac{N}{Z}\langle{}R_{n}^2\rangle{},
\end{equation}
where $\langle{}R_{p}^2\rangle{}$ and $\langle{}R_{n}^2\rangle{}$ are mean-square charge radii of the single proton and single neutron, respectively~\cite{Horowitz2012}.
In this article, $\langle{}R_{p}^2\rangle{}^{1/2}$ and $\langle{}R_{n}^2\rangle{}$ are taken as 0.84075(64)\,fm~\cite{Mohr2025} and $-0.1155(17)$\,fm$^2$~\cite{Navas2024}, respectively.
Accordingly, the parameters $c_p$ and $a_p$ were constrained such that the condition $4.6463\le\langle{}r_{p}^2\rangle{}^{1/2}\le4.6617$\,fm was always satisfied.
The red line of Fig.~\ref{fig::CS_react} shows the results of the mm-RIA calculation with the best-fit density.
The best-fit calculations are in good agreement with the experimental data.
The reduced minimum $\chi^2$ ($\chi^2_\mathrm{min}$) value, namely $\chi^2_\mathrm{min}$ divided by the number of degrees of freedom $ndf=22$, is about 1.7.

The standard errors of the parameters and radii are estimated based on the increase of $\Delta\chi^2$ corresponding to 1 standard deviation from the $\chi^2_\mathrm{min}$, expressed as 
\begin{eqnarray}
\label{eq::chi2er}
\chi^2 \le \chi^2_\mathrm{min} + \Delta\chi^2.
\end{eqnarray}
The value of $\Delta\chi^2$ obeys the $\chi^2$ probability density function for multiple parameters.
In the present case, which can be regarded as a fit with three effective parameters, $\Delta\chi^2 \approx 3.5$~\cite{Bevington2003}.
By using the Monte Carlo method, each error was evaluated from the parameter sets that satisfy Eq.~(\ref{eq::chi2er}).
The fitting results are summarized in Table~\ref{tab:densitypara}.
The error of the proton radius covers the whole range constrained by the charge radius, and the best-fit value favors the lower limit.
The rms matter radius of $^{132}$Sn is determined as $\langle{}r_{m}^2\rangle{}^{1/2}=4.758^{+0.023}_{-0.024}$\,fm.

\renewcommand{\arraystretch}{1.5}
\begin{table}
\centering
\caption{
Minimum $\chi^2$ value and best-fit 2pF parameters.
}
\label{tab:densitypara}
\begin{tabular}{ccccc}
    \hline
    $\chi^2_\mathrm{min}/ndf$ & $c_p$\,$[\mathrm{fm}]$ & $a_p$\,$[\mathrm{fm}]$ & $c_n$\,$[\mathrm{fm}]$ & $a_n$\,$[\mathrm{fm}]$  \\
    \hline
    \hline
    38.0/22 & $5.632^{+0.076}_{-0.119}$ & $0.430^{+0.063}_{-0.044}$ & $5.581^{+0.066}_{-0.068}$ & $0.576^{+0.033}_{-0.037}$ \\
    \hline
\end{tabular}    
\end{table}
\renewcommand{\arraystretch}{1.0}

\section{Comparison with theories}
The matter radius is compared with the results of \textit{ab initio} theories.
We computed the \textit{ab initio} IMSRG densities~\cite{Hergert2016} using the chiral EFT NN and three-nucleon (3N) interactions; 1.8/2.0(sim7.5)~\cite{Arthuis2024}, 1.8/2.0(EM7.5)~\cite{Arthuis2024}, $\mathrm{NNLO_{sat}}$~\cite{Ekstrom2015}, and $\mathrm{\Delta NNLO_{GO}}$~\cite{Jiang2020}.
The 1.8/2.0(sim7.5) and 1.8/2.0(EM7.5) interactions are recently developed so the binding energies and radii can be reproduced simultaneously, where the $\mathrm{NNLO_{sim}}$~\cite{Carlsson2016} and $\mathrm{N^{3}LO}$~\cite{Entem2003} NN interaction are softened by free-space similarity renormalization group evolution to the momentum resolution scale $1.8$~fm$^{-1}$, 3N interaction is regulated with a non-local regulator with the cutoff momentum $2.0$~fm$^{-1}$, and the 3N low-energy constants (LECs) are fitted to $^{3}$H and $^{16}$O binding energies and $^{16}$O charge radius, yielding 3N one-pion exchange LEC $c_{D}=7.5$ as detailed in Ref.~\cite{Arthuis2024}.
$\mathrm{NNLO_{sat}}$ includes next-to-next-to-leading order NN and 3N interaction in the expansion, and NN and 3N LECs are simultaneously optimized to data up to mass number $A=25$.
$\mathrm{\Delta{}NNLO_{GO}}$ explicitly includes the effects of $\Delta$-resonance degrees of freedom, and the LECs are simultaneously fitted to reproduce few-body data and nuclear matter properties~\cite{Jiang2020}.
All the interactions employed here can reasonably reproduce binding energies and charge radii simultaneously.
Starting from those Hamiltonians expressed within the 15 major-shell harmonic-oscillator (HO) space, we performed the Hartree-Fock (HF) calculation and rearranged the Hamiltonian based on the normal-ordered two-body approximation.
Then, we transformed the Hamiltonians with the IMSRG evolution with the two-body approximation to decouple the HF reference state from the particle-hole excited states.
The same transformation can be applied to compute the radii and densities.
The charge radii were computed with proton and neutron finite-size, Darwin-Foldy, and spin-orbit corrections.
The uncertainties are estimated from the residual HO frequency dependence, i.e.~$\hbar\omega =10$ and $12$ MeV for all interactions except $\mathrm{NNLO_{sat}}$ for which $\hbar\omega=12$ and $14$ MeV were used.

\begin{figure}[!h]
    \centering
    \includegraphics[width=0.8\linewidth]{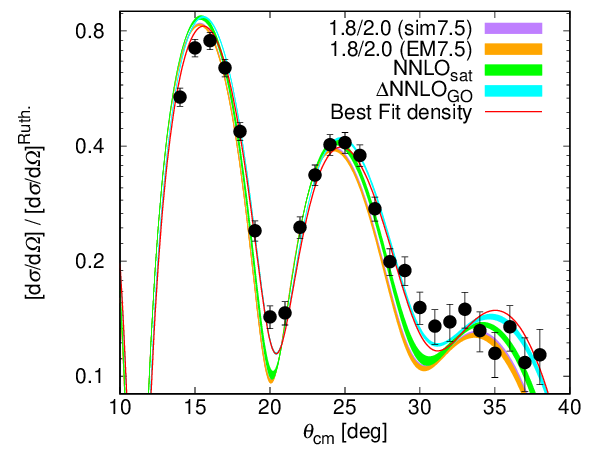}
    \caption{
    Differential cross sections divided by the Rutherford cross section.
    The red solid line shows the mm-RIA calculation with the best-fit density.
    The purple, orange, green, and cyan bands are the IMSRG calculations with the 1.8/2.0(sim7.5), 1.8/2.0(EM7.5), $\mathrm{\Delta{}NNLO_{GO}}$, and $\mathrm{NNLO_{sat}}$ interactions, respectively.
    It should be noted that the purple and orange bands nearly overlap.
    }
    \label{fig::CSRuth}
\end{figure}

Figure~\ref{fig::CSRuth} shows the Rutherford ratios of the mm-RIA calculations using the IMSRG densities, along with the experimental data.
In general, all the calculations reproduce the data ($\chi^2\sim 50\text{--}150$).
It is confirmed that the further outward the positions of the valleys around $\theta_\mathrm{cm}=30^{\circ}$, the smaller the corresponding matter radii; 4.816--4.830 (1.8/2.0 (sim7.5)), 4.816--4.830 (1.8/2.0 (EM7.5)), 4.791--4.824 ($\mathrm{NNLO_{sat}}$), and 4.736--4.746 ($\mathrm{\Delta NNLO_{GO}}$) fm.
In particular, the calculation based on $\mathrm{\Delta{}NNLO_{GO}}$ (cyan band) shows the best agreement with the experimental data and closely follows the best-fit calculation (red line).

\begin{figure}[!h]
    \centering
    \includegraphics[width=0.8\linewidth]{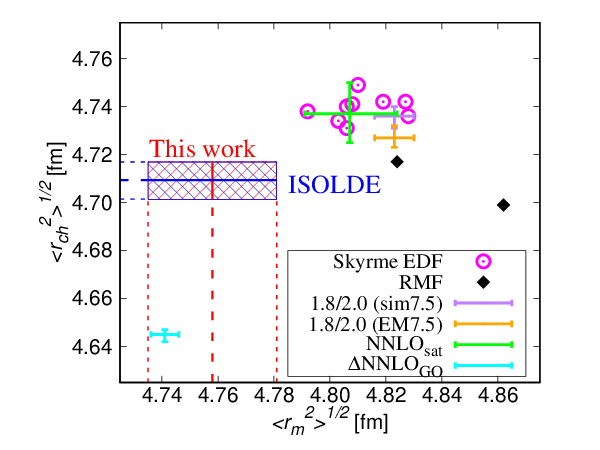}
    \caption{
    Experimental and theoretical values of the charge radius $\langle{}r_{ch}^2\rangle{}^{1/2}$ and matter radius $\langle{}r_{m}^2\rangle{}^{1/2}$ of $^{132}$Sn.
    SLy4, SLy5~\cite{SLy4SLy5}, SAMi~\cite{SAMi}, SGII~\cite{SGII}, SkM$^{*}$~\cite{SkMs}, HFB9~\cite{HFB9}, UNEDF0~\cite{UNEDF0}, UNEDF1~\cite{UNEDF1}, and UNEDF2~\cite{UNEDF2} interactions are used for the "Skyrme EDF", and the FSUGold and NL3~\cite{NL3, FSUGold, Piekarewicz2006} interactions are used for the "RMF".
    The charge radius value of the "ISOLDE" is taken from Ref.~\cite{Blanc2005}.    
    }
    \label{fig::rm}
\end{figure}

Figure~\ref{fig::rm} shows the correlation between the matter and charge radii of $^{132}$Sn predicted by theoretical models, including \textit{ab initio} methods, Skyrme energy density functionals (EDFs) with SLy4, SLy5~\cite{SLy4SLy5}, SAMi~\cite{SAMi}, SGII~\cite{SGII}, SkM$^{*}$~\cite{SkMs}, HFB9~\cite{HFB9}, UNEDF0~\cite{UNEDF0}, UNEDF1~\cite{UNEDF1}, and UNEDF2~\cite{UNEDF2} interactions, and relativistic-mean field (RMF) with FSUGold and NL3 interactions~\cite{NL3, FSUGold, Piekarewicz2006}.
The red and blue bands indicate the matter radius value of this work and the charge radius value measured at ISOLDE~\cite{Blanc2005}, respectively.
Our result implies a small radius and is consistent with the prediction from the $\mathrm{\Delta{}NNLO_{GO}}$.
However, the charge radius of the $\mathrm{\Delta{}NNLO_{GO}}$ is smaller than the experimental value.
This small matter radius seen in the present work corresponds to a small neutron skin thickness $\Delta{}r_{np}$ of approximately 0.18\,fm, suggesting a soft symmetry energy of the nuclear matter EOS~\cite{Chen2005, Cabone2010}.

\section{Summary and Outlook}
In this work, we measured, for the first time, the angular distribution of the cross sections for proton elastic scattering from $^{132}$Sn at 196--210\,MeV/nucleon over the momentum transfer ranges from 0.80 to 2.1\,fm$^{-1}$.
Using the mm-RIA calculation, we extracted the matter density distribution, where the proton and neutron density distributions were modeled with 2pF functions.
The parameters of the proton density distribution were constrained to reproduce the charge radius measured at ISOLDE.
The obtained matter radius is $4.758^{+0.023}_{-0.024}$\,fm, which is consistent with the IMSRG calculation using $\mathrm{\Delta{}NNLO_{GO}}$ interaction.
It is found that, however, combined with the charge radius measured at ISOLDE, no theoretical calculations including the IMSRG, Skyrme EDFs, and RMF calculations reproduce both experimental charge and matter radii simultaneously.
We note that a recent study quantifying three-body operator effects suggested a 1.5\% uncertainty in absolute radii for IMSRG~\cite{Heinz2025}, which is larger than the variation in the results from the NN+3N interactions examined in this work.
If a more precise many-body calculation framework is developed, a finer interaction test based on the absolute matter radii as well as charge radii will be possible.


\ack
The authors are grateful to the staff of RIKEN Nishina Center and CNS, University of Tokyo, for the operation of RIBF.
The authors would like to thank T.~Naito for the calculations of Skyrme EDFs.
This work was supported by JSPS KAKENHI Grants No.~JP15H05451, 21H04975, JP23K22500, and JP23KJ1231, and by JSPS A3 Foresight Program “Nuclear Physics in the 21st Century".
Y.~H. acknowledges support from the RIKEN Junior Research Associate Program and the JSPS Research Fellowships for Young Scientists.
P.~A.~was supported by the European Union under the Marie Skłodowska-Curie grant agreement No.~101152722. Views and opinions expressed are however those of the author(s) only and do not necessarily reflect those of the European Union or the European Research Executive Agency (REA). Neither the European Union nor the granting authority can be held responsible for them.
T.~M.~was supported by Japan Science and Technology Agency ERATO Grant No.~JPMJER2304, and JSPS KAKENHI Grant Number~25K07294.
The IMSRG calculations were performed with the imsrg++ code~\cite{Stroberg++}.
P.~A.~gratefully acknowledges the Gauss Centre for Supercomputing e.V.~(www.gauss-centre.eu) for funding this project by providing computing time through the John von Neumann Institute for Computing (NIC) on the GCS Supercomputer JUWELS at Jülich Supercomputing Centre (JSC).
Z.H.~Y acknowledges the support from the National Key R\&D Program of China (Grant No.~2023YFE0101500)


\bibliographystyle{ptephy}
\bibliography{ptep}


\vspace{0.2cm}
\noindent


\let\doi\relax


\end{document}